\documentclass[twoside,leqno,twocolumn]{article}

\usepackage[letterpaper]{geometry}
\usepackage{graphics}
\usepackage{graphicx}
\usepackage{ltexpprt}
\usepackage{xcolor}
\newcommand{\citep}[1]{\cite{#1}}
\newcommand{\citet}[1]{\cite{#1}}

\begin{document}

\title{\Large Load-Balanced Bottleneck Objectives in Process Mapping\thanks{This work was partially supported by the Austrian Science Foundation~(FWF, project P 31763-N31),
and it was partially supported by the Research Council of Norway (project RCN 251186).}}

\author{Johannes Langguth\thanks{Simula Research Laboratory, Oslo, Norway}
\and Sebastian Schlag\thanks{Karlsruhe Institute of Technology, Karlsruhe, Germany}
\and Christian Schulz\thanks{University of Vienna, Faculty of Computer Science, Vienna, Austria}}

\date{}

\maketitle







\begin{abstract} \small\baselineskip=9pt We propose a new problem formulation for graph partitioning that is tailored to the needs of time-critical simulations on modern heterogeneous supercomputers. \end{abstract}

\section{Introduction}
Among the many combinatorial problems, graph partitioning (GP) has a central role in the area of parallel high performance computing. Irregular inputs such as graphs, sparse matrices and many meshes typically have to be distributed among the nodes of a supercomputer to allow parallel processing, and the distribution should both balance the workload and minimize the communication. In the last two decades, sophisticated software has been developed for computing high-quality partitionings fast. However, the underlying model, which assumes homogeneous nodes connected by a full-bisection bandwidth network, no longer represents current supercomputers. We therefore propose a new problem formulation which overcomes these drawbacks and discuss its suitability for future challenges in parallel computing. 

Irregular inputs can roughly be divided into two main groups. The first consists of problems whose data access pattern resembles sparse-matrix-dense-vector multiplications (SpMV), where the communication is irregular but follows a repeated pattern, such as PageRank \cite{ilprints422} and many scientific computations on irregular meshes. The second group represents problems resembling sparse-matrix-sparse-vector multiplications (SpMSpV). Here, only a subset of vertices is active at a time, leading to changing communication patterns. While minimizing cut sizes helps to reduce the communication cost in both cases, in the first case, the overall communication time is determined by the maximum communication per node, assuming the network links of a distributed memory computer operate independently. Thus, minimizing the maximum
(i.e.,~the makespan) is a better partitioning objective, and assuming there are no slowdowns such as concurrent traffic in the network, it is actually optimal in this case. In the second case, it is harder to compare the partitioning objectives. In low-diameter graphs such as social networks, breadth-first search and similar algorithms typically perform a low number of high-volume communication rounds. Consequently, minimizing the maximum communication among the nodes may lead to better results than cut-size optimization. For high diameter graphs this no longer holds, which renders makespan partitioning less attractive in this case. 

Finally, current single-criterion partitioning does not allow trade-offs between load balance and cut size. For SPMV type computations, where both communication and computation performance is typically to a certain degree predictable, incorporating this trade-off has the potential to provide better solutions than using a fixed load-balance constraint on the vertices. And while SpMSpV computations are typically less predictable, they are often highly communication-bound, which means that load-balance is less important than communication minimization and might benefit from the trade-off formulation. All this however implies that we must provide a ratio between the cost of communication and computation for the trade-off to function.

\section{Related Work}
There has been an enormous amount of research on GP, and we refer
the reader to Refs.~\cite{GPOverviewBook,SPPGPOverviewPaper,DBLP:reference/bdt/0003S19} for extensive
material and references. 
The most common objective function is to minimize the total cut size $\sum_{i<j}\omega(E_{ij})$. 
There are well known software packages based to find partitions minimizing this objective including KaHIP~\cite{kaHIPHomePage}, Jostle~\cite{Walshaw07}, Metis \cite{KarypisK98a}, and Scotch \cite{Scotch}. 
A related objective is to minimize the \emph{maximum cut size} $\max_{i<j}{\omega(E_{ij})}$.
The first accurate communication volume metrics for sparse-matrix vector multiplication, using hypergraph models, are due to Catalyurek and Aykanat~\cite{DBLP:journals/tpds/CatalyurekA99}. 
When \emph{graph} partitioning is used in parallel computing to map the graph nodes to different processors, the \emph{communication volume} is often more representative than the cut~\citep{hendrickson2000graph}.
Let $V_\ell$ be the block containing vertex $v$. Then, we let $D(v)$ denote the number of blocks in which vertex $v$ has a neighbor, excluding the block containing $v$ and $c(v)$ be a vertex weight. That is, $D(v) = | \{ j \mid \exists u \in V_j \neq V_{\ell} \textrm{\ s.t.\ } \{u,v\}\in E \} |$. For a block $V_i$, the \emph{communication volume} is $cvol(V_i) :=  \sum_{v\in V_i} c(v)D(v)$. Similar to cut size, one can minimize the \emph{maximum communication volume} $\max_{1\leq i \leq k} cvol(V_i)$, or the \emph{total communication volume} $\sum_{1\leq i\leq k} cvol(V_i)$. The choice of either the total or maximum variant of the objective function here is typically dictated by the topology of the interconnection network between the machines~\citep{hendrickson2000graph}.
Other objectives are less common, these include the maximum degree in the \emph{quotient graph} (the graph formed by blocks and their connections), expansion and conductance~\citep{lang2004flow}, and optimizing the ``shape'' of partitions~\citep{diffusive-2012}. Specialty partitioners tailor-made for applications can give better quality~\citep{natural-cuts-2011}. Furthermore, there are methods for maximizing multiple objective functions simultaneously~\citep{Schloegel1999,subdomain-degree-minimization-2003,umpadimacs}, and finding Pareto-optimal solutions~\citep{hamann2015graph}. 

There is likewise a large literature on process mapping, i.e., decomposition techniques that takes the given connection network into account, often in the context of scientific applications using MPI (Message-Passing
Interface).  
Refs.~\cite{DBLP:conf/irregular/WalshawCEJM95,DBLP:journals/pc/SadayappanER90,catalyurek1995hypergraph,Hatazaki98} were among the first to tackle the process mapping problem. 
Hatazaki~\cite{Hatazaki98} and Walshaw~\cite{DBLP:conf/irregular/WalshawCEJM95} proposed graph partitioning to solve the MPI process mapping problem for unweighted process topologies.  Mercier and
Clet-Ortega and later
Jeannot~\cite{mercier2009towards,MercierJeannot11} simplify the
mapping problem by only considering the topology inside the compute
nodes themselves and ignoring the topology of the network. Multiple
placement policies are investigated to enhance overall system
performance.  Yu et al.~\cite{YuChungMoreira06} discuss and implement
embedding heuristics for the BlueGene $3d$ torus system.  Hoefler and
Snir~\cite{HoeflerSnir11} instead optimize the congestion of the
mapping and show that this problem is NP-complete. Routing-aware mapping
heuristics taking the hierarchy of specific hardware topologies into
account were discussed in Ref.~\cite{Abdel-GawadThottethodiBhatele14}.
Later, based on a variety of techniques, more sophisticated algorithms to compute mappings for hierarchical systems have been published~\cite{schulz2017better,DBLP:conf/icpp/GlantzPM18,fonseca2020better}.

The \textit{Lynx} code \cite{micro} is an SpMV-type computation for large scale simulations in cardiac electrophysiology on heterogeneous supercomputers. As the problem is relatively communication heavy, high-quality partitioning is crucial for its performance, especially when using nodes equipped with multiple GPUs. Scalability was achieved via hierarchical partitioning. However, due to the lack of actual hierarchical partitioning software, it was emulated by applying conventional partitioning twice. This proved to be highly effective, but difficult to program. 
As this type of compute node has become common in modern supercomputers, we believe that the need for hierarchical partitioning software will increase in the~near~future.

\section{Problem Specification}
We define the basic version of the \textit{graph-constrained makespan partitioning problem} as follows:

Given a tree $C=(B,L)$, an undirected graph $G=(V,E)$, and a communication cost factor $F$. Find a partitioning $P: V \rightarrow B$ which maps each vertex $v \in V$ to a vertex $b \in B$ such that the \textit{makespan} is minimized. We refer to the vertices $B$ of $C$ as \textit{bins}, and the edges $L$ as links. The \textit{makespan} $M$ is defined as:
$$M(P)=\max(\max_{b \in B}comp(b),F\cdot\max_{l \in L}comm(l))$$

where $comp(b)=|\{v \in V \mid P(v)=b\}|$ is the computational load of vertex $b \in B$, $comm(l)= |\{\{u,v\} \in E \mid l \in STP(P(u),P(v)) \}|$
is the communication volume along edge $l \in L$, and $STP(P(u),P(v))$ is the shortest path (as a set of edges) between the bins $P(u)$ and $P(v)$ that the endpoints $u$ and $v$ of some edge were assigned to in $P$.
Because $C$ is a tree, the shortest path is unique.
Note that $C$ is unweighted, which means that the length of a path is simply the number of edges contained in it. This is based on the intuition that if an edge is cut, communication will be necessary between the compute nodes represented by these bins.

\subsection{Generalizations.}
The most basic generalization is the introduction of a set $R \subset B$ of routers, i.e., bins $r \in R$ that
cannot be assigned any work (i.e., $load(r)=0$ $\forall r \in R$). This is needed to correctly model the networks of most supercomputers. We refer to this variant as the \textit{interconnect graph-constrained makespan partitioning problem}.

To take more complicated interconnects with different routing protocols into account, the protocols become part of the problem formulation. Thus, the \textit{routing graph-constrained partitioning problem} differs from the standard version in that $C$ is no longer required to be a tree. Any algorithm for this problem is given access to an \textit{oracle} (such as a routing table) which for every pair of
bins returns a unique path between them. If we allow for multipath routing, the oracle may instead return a set of paths. In that case, assuming $k$ paths, each path $\mathcal{P}$ adds $1/k$ to the load of each edge $l \in \mathcal{P}$.
 
The hierarchical network structure of interconnect topologies such as fat trees implies that some links need to have much higher capacities than the injection bandwidth of a single node. In order to model this, we change the communication factor $F$ from globally applying to all links to a link-specific factor $F_l$, thereby obtaining the \textit{edge-weighted} variant. By the same token, we can define a \textit{vertex-weighted} variant where every $v \in 
V$ is given a weight $w(v)$. The load of a bin is then computed as the sum of the weights of all vertices assigned to that bin.

\subsection{NP Hardness.}
Unlike the traditional graph partitioning problem, graph constrained makespan partitioning does not have a simple reduction from \textsc{Minimum Bisection}. However, the vertex weighted case can easily be proven to be NP-Hard by reduction from \textsc{Minimum multiprocessor scheduling} \cite{GareyJohnson}.

\bibliographystyle{plain}
\bibliography{quellen,phdthesiscs,applications,oldbib,lecture}

\end{document}